\documentclass[12pt]{article}
\usepackage{subeqnarray,eqsection,indent,cite}
\usepackage{latexsym}
\usepackage{amsmath}           
\usepackage{amsfonts}          
\usepackage{amssymb}           
\usepackage{mathrsfs}

\newcommand{\eMut}{e^{M_{t+1} }U_{0,t}^\dag}
\newcommand{\eMvt}{U_{0,t} e^{M_{t+1}}}
\newcommand{\emenMut}{U_{0,t} e^{-M_{t+1}} }
\newcommand{\emenMvt}{ e^{-M_{t+1}}U_{0,t}^\dag}

\newcommand{\eMvtmenone}{U_{0,t-1} e^{M_{t}}}

\newcommand{\meta}{\frac{ 1}{2}\,}

\newcommand{\uh}{{\hat u}}
\newcommand{\vh}{{\hat v}}
\newcommand{\Iddop}{{\mathbb I}{\mathrm d}}
\newcommand{\ut}{{\tilde u}}

\newcommand{\bone}{1\!\!1}

\newcommand{\alphah}{{\hat \alpha}}
\newcommand{\betah}{{\hat \beta}}

\newcommand{\ket}[1]{\vert #1 \rangle}
\newcommand{\bra}[1]{\langle #1 \vert}
\newcommand{\braket}[2]{\langle #1 \vert #2 \rangle}

\renewcommand{\det}{\mathrm{det}\:}
\newcommand{\tr}{\mathrm{tr}}
\newcommand{\Tr}{\mathrm{Tr}}
\newcommand{\tg}{\mathrm{tg}}
\newcommand{\misuradif}{\mathrm D}
\newcommand{\Fifield} {\phi}  
\newcommand{\fc}{{\cal F}}

\newcommand{\giro}[1]{\stackrel{\circ}{#1}}
\newcommand{\nc}{\giro{R}}
\newcommand{\mc}{R}
\newcommand{\nt}{\nc\hspace{-1mm}{}_t}
\newcommand{\mt}{\mc_t}
\newcommand{\me}{E}
\newcommand{\md}{\giro{E}}
\newcommand{\Laa}{L}          
\newcommand{\Lbb}{\giro{L}}  
\newcommand{\Iaa}{I}           

\newcommand{\BNd}{\giro{\cal F}_{N,\,t}}    
\newcommand{\NdB}{{\cal F}_{N,\,t}  }   
\newcommand{\BNdt}{\giro{\cal F}_{N,\,t+1}}   
\newcommand{\NdBt}{{\cal F}_{N,\,t+1}}  
\newcommand{\BNdtm}{\giro{\cal F}_{N,\,t-1}  } 

\newcommand{\pezzo}{\sigma }

\newcommand{\be}{\begin{equation}}
\newcommand{\ee}{\end{equation}}

\newcommand{\reff}[1]{(\ref{#1})}

\begin{document}

\title{ Bogoliubov transformations and fermion condensates in
lattice field theories}


\author{
  {\small Sergio Caracciolo}                       \\[-1.7mm]
  {\small\it Dipartimento di Fisica and INFN}      \\[-1.7mm]
  {\small\it Universit\`a degli Studi di Milano}   \\[-1.7mm]
  {\small\it via Celoria 16, I-20133 Milano, ITALY}                \\[-1.7mm]
  {\small\tt Sergio.Caracciolo@mi.infn.it},        \\[-1.7mm]
  {\protect\makebox[5in]{\quad}}
   \\
  {\small Fabrizio Palumbo}  \\[-1.7mm]
  {\small\it INFN -- Laboratori Nazionali di Frascati}      \\[-1.7mm]
  {\small\it P.~O.~Box 13, I-00044 Frascati, ITALY}       \\[-1.7mm]
   {\small\tt fabrizio.palumbo@lnf.infn.it}       \\[-1.7mm]
  \\
  {\small Giovanni Viola}  \\[-1.7mm]
  {\small\it Dipartimento di Fisica and INFN}      \\[-1.7mm]
  {\small\it Universit\`a degli Studi di Firenze}       \\[-1.7mm]
  {\small\it via G. Sansone 1, I-50019 Sesto Fiorentino Firenze, ITALY}                \\[-1.7mm]
   {\small\tt giovanni.viola@fi.infn.it}       \\[-1.7mm]
 {\protect\makebox[5in]{\quad}}  
}



\maketitle

\thispagestyle{empty}

\begin{abstract}

We apply generalized Bogoliubov transformations to the transfer matrix of relativistic field theories regularized
on a lattice. We derive the conditions these transformations must satisfy to factorize the
transfer matrix into  two terms which propagate fermions and antifermions separately, and we solve the
relative equations under some conditions. We relate these equations to the saddle point approximation
of a recent bosonization method and to the Foldy-Wouthuysen transformations which separate positive
from negative energy 
states in the Dirac Hamiltonian.


\end{abstract}

\clearpage

\vfill\eject

\renewcommand{\thefootnote}{\arabic{footnote}}
\setcounter{footnote}{0}

\clearpage


\section{Introduction}

In this paper we investigate some properties of Bogoliubov transformations in fermionic lattice field theories in connection
with the appearance of  fermionic  condensates. 
We are motivated by our study of a new bosonization method~\cite{Cara}, but our results have a wider relevance, and
relate these transformations to the ones introduced by Foldy-Wouthuysen to separate fermions from
antifermions in the Dirac Hamiltonian~\cite{FW}.

Bogoliubov transformations  are unitary transformations which mix creation and annihilation operators. They have been introduced in the theory of many-body systems in which have been extensively used for their simplicity, in particular in connection with variational principles, and also as a starting point of more elaborated approximation methods. 
In their application to superconductivity they reproduce the results of the BCS theory of
electron-electron interactions with an easy extension to the case of electrons interacting with phonons~\cite{Blatt}. 

While mixing of creation-annihilation operators was a novelty in the theory of many-body systems,  mixing
of particles and antiparticles in the Hamiltonian formalism of relativistic field theories is very natural, and 
in fact  generalized Bogoliubov transformations have been used also in this domain. See for example~\cite{Araki} where general bilinear fermionic Hamiltonians are diagonalized, or more recents studies of QCD in the limit of large number of colours~\cite{Bardeen}. Bogoliubov transformations are a fundamental tool in understanding the black body radiation in the Unruh effect in accelerating reference frame~\cite{Unruth} and the black hole Hawking radiation~\cite{Wald}. But the specific difficulties of the renormalization procedure in the Hamiltonanian formalism have limited their use in quantum field theory.

Recently Bogoliubov transformations have found a somewhat different application in lattice field theory in the formalism of the transfer matrix, 
which is close from the physical point of view to the Hamiltonian formalism.
 They have been used  to introduce dynamical composite bosons in  fermionic theories.  An independent  Bogoliubov transformation at each time slice can be performed in the operator form of the partition function.
The time-dependent parameters of the transformation are then associated with composite bosonic fields in the presence
of fermionic fields (quasi-particles) satisfying a compositeness condition which avoids double counting~\cite{Palu, Palub}. One thus gets an effective action of composite fields plus quasi-particles, exactly equivalent to the original one, in which ground state
and excited states can be treated on the same footing.  
Of course, in practical applications, some approximation must be introduced.

A Bogoliubov transformation generates a new vacuum which has the form of a fermion condensate.  We 
studied such a condensate in  a first  approach to bosonization~\cite{Cara}, which can be regarded as an approximation of the method of Ref.~\cite{Palu, Palub}, in which quasi-particles are altogether neglected.  In an application to a four-fermion interaction model with a discrete chiral symmetry (at zero fermion mass), an explicit form of the fermion condensate appearing when the symmetry is spontaneously broken was found in a saddle point approximation. 
The condensate is made of a composite boson  which is a superposition of a symmetry breaking plus a symmetry conserving state. This result is not surprising from the point of view of the renormalization group, because 
it tells that we have a contribution of two operators of the same dimension which are no longer separated by a symmetry.
But looking carefully at this result we found that condensation of a symmetry conserving boson takes place also
in the free theory, if we require factorization of the transfer matrix in two terms which propagate
particles and antiparticles separately. This requirement gives rise to the  same equations as the
requirement  of extremality of the vacuum energy which is generated by the Bogoliubov transformation,
in the same way as in the many-body theory~\cite{Blatt}. 

The saddle point approximation of the effective action we are talking about, under the assumption that the saddle point equations have stationary solutions,  equals  the effective action obtained after a time independent Bogoliubov transformation by neglecting quasi-particles, without any reference to the bosonization method. We will present our results in this more general perspective,  but keeping in mind that,  by  time
independent Bogoliubov transformations, we are investigating the saddle point stationary solutions of the mentioned
approximated bosonization method.

At this point the relation of Bogoliubov transformations with Foldy-Wouthuysen transformations should be clear. The latter ones eliminate the mixing between positive and negative energy solutions in the continuum free Dirac Hamiltonian.
 The explicit form of these transformations  has also been found for some interactions, which include minimal~\cite{Case} and anomalous~\cite{EK} interaction of spin-$1/2$ particles with a time-independent magnetic field, anomalous interactions with a time-independent electric field and with a pseudo-scalar field~\cite{MMZ}. Other generalizations are discussed in~\cite{Nik}. See also~\cite{Sil} and references therein.

In the second quantization formalism the separation of  positive- from negative-energy states is accompanied by the generation of a new vacuum.  Of course in the free-field case this vacuum will be unitary-equivalent to the previous one, but more intricate possibilities are opened in the interacting case in presence of a phase transition. We found also interesting to see how, in the correspondence from the first and second quantization, the ambiguities in the Foldy-Wouthuysen transformation~\cite{EK} appear as multiple solutions of the saddle-point equations for the corresponding Bogoliubov transformation, and they are solved by demanding that the new vacuum energy be minimal.

In this paper we  study these transformations and the corresponding fermi\-onic condensates for relativistic field theories regularized on a space-time lattice which are suitable for non-perturbative studies and numerical simulations.
The elimination of a mixing between fermions and antifermions corresponds to a factorization 
of the transfer matrix. One then wonders which transformation, at finite lattice spacing, induces this 
factorization and how it changes the properties of the vacuum which appears, 
in general, as a condensate of pairs of the original fermion-antifermions. 
One way is of course by studying the implementation 
of symmetries, but for a symmetry which is broken by the regularization, 
like the chiral one with Wilson fermions, for instance, this can present some difficulty. 
 
 The paper is organized in the following way. In Section 2 we define the generalized Bogoliubov transformations
 and derive the effective action, extending the result of ~\cite{Palub} in which quasi-antiparticles were neglected. In
 this effective action there appear quadratic terms mixing quasi-particles and quasi-antiparticles and a 
 vacuum energy which has the form of a condensate of a composite boson. In Section 3 we derive and solve, at finite lattice spacing,  the decoupling equations. They are  identical to the ones arising from the requirement of extremality of the vacuum energy,  under some simplifying hypotesis. In Section 4 we consider the properties of the condensate in connection with chiral invariance. In Section 5 we discuss 
 the relation with the Foldy-Wouthuysen transformation. In Section 6 we write the effective action at
 finite temperature in terms the ``physical" modes and in Section 7 we present our Conclusions. Many details of
 our derivations are relegated to the Appendices.

\section{Generalized Bogoliubov transformation}

Consider a system of fermions interacting  with  external bosonic fields 
including gauge fields, regularized on a lattice. The fermionic part of the partition
function at finite temperature $T$ and chemical potential $\mu$ can be written
\begin{equation}
\mathcal{Z} = \mbox{Tr}^{F}  
\prod_{t=0}^{L_0/s-1}  e^{M_t + M_t^{\dagger}} \,
{\cal T}_{t,t+1}
\label{part}
\end{equation}
where $L_0 =  T^{-1}$  is the number of links in the temporal direction, $\cal T$
is the fermion transfer matrix and  $\mbox{Tr}^F$ is the trace  over the Fock space of fermions. The parameter $s$ takes the value $1$ in the Wilson formulation for lattice fermions, but $s=2$ for the Kogut-Susskind fermions which live on blocks of size twice the lattice spacing. The index $t$ labels  the blocks along the ``time" direction.
The 
expression of $\cal T$ which we will use was given by Lus\"cher~\cite{Lusc}, for Wilson fermions, in the gauge
 $U_0 = \bone$,
in which  one has to impose the Gauss constraint in the Hilbert space of the system (a Fock space of fermions in which the coefficients of the fermionic states are  polynomials of 
spatial link variables). Here we report
a slightly modified form which avoids the Gauss constraint by reinstating the temporal links variables which generate the Gauss constraint
 \be
{\cal T}_{t,t+1} :=  \hat{T}^\dagger_t  \, {\hat V}_t \exp(s\,\mu \, \hat{n}) \hat{T}_{t+1}
\ee
where $\hat{n}$ is the fermion number operator and the sum on all the indices is understood
\be
\hat{n} :=  \hat{u}^{\dagger} \hat{u} - \hat{v}^{\dagger} \hat{v} \, ,
\ee
with $\hat{u}^\dagger$ and $\hat{v}^\dagger$, creation 
operators of fermions and antifermions, obeying canonical anti-commutation relations and 
\begin{eqnarray}
\hat{T}_t  &=&\exp [ -\hat{u}^{\dagger} M_t\, \hat{u} - \hat{v}^{\dagger}  
M^T_t \hat{v} ] 
\exp[\hat{v} N_t  \, \hat{u}] \\
{\hat V}_t &=& \exp  [ \hat{u}^{\dagger}\ln  U_{0,t}\, \hat{u} + \hat{v}^{\dagger}  \ln
U_{0,t}^* \, \hat{v} ] \,.
\end{eqnarray}
The matrices $M_{t}$ ($M_t^T$ being the transposed of $M_t$) and $N_t$ are functions of the  spatial 
link variables at time $t$ and possibly of other  bosonic fields.
Explicit expressions for  Wilson and Kogut-Susskind fermions in the flavor basis are  reported 
in  Appendix~\ref{transfer}.
The variables $ U_{0,t}$ are matrices in a unitary representation of the gauge group whose  elements are the link variables between Euclidean time $t$ and $t+1$
\begin{equation}
(  U_{0,t} )_{{ \bf x}_1,{ \bf x}_2 }
= \delta_{{ \bf x}_1,{ \bf x}_2 } U_{0,t} ( {\bf x}_1)
\end{equation}
where boldface letters,
as $\mathbf{x}$,  denote spatial coordinates.

We introduced the notation, which we will use  for any matrix $\Lambda$
\begin{equation}
\mbox{tr}_{\pm} \Lambda := \mbox{tr}\left( P_{\pm} \Lambda \right) \,.
\end{equation}
The  operators $P_{\pm}$  project on the  components of the fermion field which propagate forward or backward in time
\begin{eqnarray}
u &=& P_{+} \psi
\nonumber\\
v^{\dagger} &= &P_{-} \psi
\end{eqnarray}
and their expressions are given in the Appendix~\ref{transfer}. The symbol
 ``$\mbox{tr}$" denotes  the trace over fermion-antifermion intrinsic quantum numbers and spatial coordinates (but not over time).

Let us compare the transfer matrix to the Hamiltonian
\be
H = M  \, \hat {c}^{\dagger} \hat{c} +N ( \hat {c}^{\dagger} \hat{c}^{\dagger}  +
\hat {c} \, \hat{c})\,.
\ee
This Hamiltonian, which contains only one fermion operator, can be diagonalized by a Bogoliubov transformation. We then try to factorize the transfer matrix by its natural generalizion. However, we have the transfer matrix in the form of a product of non-commuting operators. Even in the simplest case when $M_t= \ln U_{0,t}= 0$, so that
\be
{\cal T}_{t,t+1} = \exp[\hat{u}^\dag N_t^\dag  \, \hat{v}^\dag] \exp[\hat{v} N_{t+1}  \, \hat{u}]
\ee
it is not possible to solve the problem by diagonalization of each factor. Indeed, $\hat{v} N_{t+1}  \, \hat{u}$ cannot be diagonalized. To diagonalize the product the strategy is, also in this more complex situation,  to perform a generalized Bogoliubov transformation which generates new creation-annihilation  operators of ``physical" particles  
\begin{subeqnarray}
{\hat \alpha}  =  R^{\frac{1}{2}}\left( {\hat u} -  
 \,{\mathcal F}^{\dagger}  \, {\hat v}^{\dagger}\right) 
&\qquad&
{\hat \beta}  =  \left( {\hat v} +  
{\hat u}^{\dagger} \,{\mathcal F}^{\dagger} \right) \giro{R}{\hspace{-1mm}}^{\frac{1}{2}} \label{Bogoliubov} \\
{\hat \alpha}^{\dagger}  =  
	\left( {\hat u}^{\dagger} - {\hat v} \, {\mathcal F}
	\right)  R^{\frac{1}{2} }  &\qquad& 
{\hat \beta}^{\dagger}  =  \giro{R}{\hspace{-1mm}}^{\frac{1}{2} } 
\left( {\hat v}^{\dagger} + {\mathcal F}\,
{\hat u}\right) 
\end{subeqnarray}
where
\begin{equation}
 R= (1 + {\mathcal F}^{\dagger} {\mathcal F})^{-1} \qquad  \giro{R} = (1 + {\mathcal F} {\mathcal F}^{\dagger})^{-1} 
\label{involution}
\end{equation}
and $ {\mathcal F}$ is an arbitrary matrix. The upperscript circle denotes the involution
defined by the above equations.  The new operators satisfy 
 canonical commutation relations for any choice of the matrix $ {\mathcal F}$. We will let $ {\mathcal F}$ to depend on all the fields coupled to the fermions in such a way as to respect all symmetries. The vacuum of the new operators is
 \be
  |\mathcal{F} \rangle= \exp \hat{\mathcal{F}}^{\dagger}|0 \rangle \label{defF}
  \ee
  where
\begin{equation}
{\hat \fc }^{\dagger} = \uh^\dag \fc^\dag  \vh^\dag,
\end{equation}
is a the creation operator of a composite boson. As already said the new vacuum appears as a coherent state of  fermion-antifermion pairs.

Usually  the trace appearing in the definition of the transfer matrix is evaluated using coherent states of fermions
\begin{equation}\label{coherent}
| \alpha,\beta\rangle = \exp (- \alpha  {\hat u}^{\dagger} - \beta {\hat v}^{\dagger} ) | 0 \rangle,
\end{equation}
where the $\alpha ,\beta$ are Grassmann fields. We will use instead states
obtained by applying a Bogoliubov transformation  
\begin{align}\label{bogcoherent}
 \hat{\mathscr U}(\mathcal{F})\, | \alpha,\beta\rangle = & | \alpha,\beta; \fc\rangle  \\
 = & \exp (- \alpha \, {\hat \alpha}^{\dagger} - \beta \,{\hat \beta}^{\dagger} ) |\mathcal{F} \rangle \\
= & \exp \left( \uh^\dag {\cal F }^\dag \vh^\dag - a\,   \alphah^\dag - b\,  \betah^\dag - \beta {\cal F } \alpha \right) \big|0 \rangle 
\end{align}
where $ |\mathcal{F} \rangle$ is the coherent state of fermion-antifermion pairs defined in~\reff{defF}, $a:=\mc^{-\meta}\alpha$ and $b:=\beta \giro{R}{\hspace{-1mm}}^{-\frac{1}{2}}$. The explicit definition of the operator $ \hat{\mathscr U}$ is given in Appendix~\ref{transformation}.

In terms of the transformed coherent states the partition function can be written 
\begin{equation}
Z= \mbox{Tr}^{\mbox{F}} \prod^{L_0/s-1}_{t=0}\Big[ e^{M_t + M_t^{\dagger}}  {\hat {\mathcal P}} \, \hat T_t^\dag\hat V_t  \, e^{s\,\mu \, {\hat n}} \,\hat T_{t+1}\Big]\\
\end{equation}
where
\begin{equation}
 {\hat {\mathcal P}} :=\int{\mathrm  D}[\alpha^*,\alpha,\beta^*,\beta]  \frac{\ket{ \alpha \beta;\fc }\bra{ \alpha \beta;\fc }}{\braket{  \alpha \beta;\fc}{ \alpha \beta;\fc} } 
\end{equation}
 is a representation of the identity. More explicitly
\be
Z = \int \mathrm{D} [\alpha^*,\alpha,\beta^*,\beta] \, \prod_t\,e^{M_t + M_t^{\dagger}} \frac{ \bra{\alpha_t \beta_t ; \fc _t}\hat {T}_t^{\dag} \hat {V}_te^{s \mu \hat n}\hat {T}_{t+1}\ket{ \alpha_{t+1} \beta_{t+1};\fc _{t+1}}   }{\braket{ \alpha_t \beta_t;\fc _t}{
\alpha_t\beta_t;\fc _t }^{-1}}
\ee
where the Grassmann variables $\alpha^*, \alpha, \beta^*, \beta$ satisfy antiperiodic boundary conditions in time.
Evaluating the trace as outlined in the Appendix~\ref{trace} we get
\begin{eqnarray}\label{Z}
Z &=
\int{D[\alpha^*,\alpha,\beta^*,\beta]\, e^{-S_0 (\fc)-S_F(\alpha,\beta;\fc) }}\,.
\end{eqnarray}
In the above equation $S_0 (\fc)$ is a term independent of the Grassmann variables 
which can be interpreted as a vacuum energy
\be\label{bosonaction}
   S_0(\fc):= - \, \sum_{t=0}^{L_0/s-1}\tr_{+} \ln\left ( R_t \,U_{0,t}\, \me_{t+1,t}  \right)
\ee
where
\be
 \me_{t+1,t} :=   \,\left(\NdBt\right)^\dag \eMut \,e^{M_t^\dag}\,\NdB
 +\, \fc_{t+1}^\dag\,\emenMvt \, e^{-M_t^\dag}\, \fc_{t}\,,  
\ee
with
\be
 \NdB :=1+N^{\dag}_{t} \fc_{t}\,.
\ee
The other term is the action of quasi-particles 
\begin{multline}\label{SF}
S_F(\alpha,\beta;\fc)=  - s \sum_{t=0}^{L_0/s-1}\Big[ \beta_{t} {I }_t^{(2,1)} \alpha_{t} + \alpha^*_{t}{I }_t^{(1,2)} \beta^*_{t}\\
	 + \alpha^*_t  (\nabla_t -{\cal H}_t) \alpha_{t+1}-\beta_{t+1}  (\giro{\nabla}_t-\giro{{\cal H}_t)}\beta^*_{t}\Big]  
\end{multline}
 written in terms of lattice covariant derivatives
\begin{align}
\nabla_t&:=s^{-1}\left(e^{s\mu}U_{0,t}-T^{(-)}_0\right) \\
\giro{\nabla}_t&:=s^{-1}\left( e^{-s\mu}U_{0,t}^\dag -T^{(+)}_0\right)
\end{align}
and fermion-antifermion Hamiltonians which are given by
\begin{align}
{\cal H}_t \,:= &\, s^{-1}e^{s\mu}\left(  U_{0,t} - \mt^{-\frac{1}{2}}\me_{t+1,t}^{-1}\mc_{t+1}^{-\frac{1}{2}} \right) \\
\giro{{\cal H}}_t\, :=& \,  s^{-1}e^{-s\mu}\left( U_{0,t}^\dag - \nc\hspace{-1mm}{}^{-\frac{1}{2}}_{t+1} \giro{\me}\hspace{-1mm}{}_{t+1,t}^{-1}\nt^{-\frac{1}{2}}
\right)  
\end{align} 
plus the unwanted terms which mix  quasi-particles with quasi-antiparticles whose coefficients are
\begin{align} 
  \Iaa_t^{(2,1)} & := 
 \, s^{-1}\,  \nt^{\frac{1}{2}}\,  \left[ \nt-\giro{\me}\hspace{-1mm}{}_{t,t-1}^{-1}\BNdtm ~e^{M_{t-1}^{\dagger}}\eMvtmenone  \right] \fc_{t}^{\dag -1}  \mc_{t}^{\frac{1}{2}} \\
 \Iaa_t^{(1,2)} &:= 
  \, s^{-1}  \mc_{t}^{\frac{1}{2}}\,  \fc_{t}^{-1}\left[ \nt-  e^{M_{t}^\dag}~\eMvt~\left(\BNdt\right)^\dag \giro{\me}\hspace{-1mm}{}_{t+1,t}^{-1}\right]\, \nt^{\frac{1}{2}}
\end{align}
$T_0^{(\pm)}$ are the forward and backward
translation operators of one block, that is $s$ lattice spacing, in the ``time"  direction 
\begin{equation}
[T_0^{(\pm)}]_{t_1,t_2 }=\delta_{t_2,t_1 \pm 1}
\end{equation}
and the definitions of the other new symbols are
\begin{align}
 \giro{\me}_{t+1,t}\,  := &  \,  \BNd \,e^{M_t^\dag}\,\eMvt\,\left(\BNdt\right)^\dag
	  +\,  \fc_{t}\,e^{-M_t^\dag}\,\emenMut\fc_{t+1}^\dag \\
  \BNd \,  := &  \, 1+\fc_{t} N^{\dag}_{t}\, .
\end{align}

\section{Factorization of the transfer matrix}

In the study of the factorization of the transfer matrix we will consider both fermion regularizations, but since 
the main motivation is to disentangle unphysical condensates from condensates related
to symmetry breaking,  the Kogut-Susskind regularization is more interesting because it conserves a form of
chiral invariance.

The  fermion-antifermion mixing can be eliminated whenever  $\cal F$ can be chosen in such a way that the equations 
\be
 		\Iaa{}_t^{(2,1)}\,=\,\Iaa{}_t^{(1,2)}=0
\ee
are satisfied. Therefore for these particular values of  $\cal F$ the transformations~\reff{Bogoliubov} factorize the transfer matrix.

It is remarkable that the requirement that the vacuum energy be minimal (namely  extremality
of the vacuum energy with respect to ${\mathcal F},{\mathcal F}^{\dagger} $) gives rise to the same equations, in full analogy with the Bogoliubov transformation in the BCS theory~\cite{Blatt}.

Explicitly  we get
\begin{align}
\fc_{t+1}\,= \,& N_{t+1} +\emenMvt  e^{-M_t^\dag} \fc_t \big(\fc_{N,\,t}\big)^{-1} e^{-M_t^\dag} \emenMut  \\
 \fc_{t}^\dag\,= \,& N_{t}^\dag + e^{-M_t^\dag}\emenMut  \big(\fc_{N,\,t+1}^\dag\big)^{-1} \fc_{t+1}^\dag \emenMvt e^{-M_t^\dag}\,.
\end{align}
We notice that the form of these equations does not depend  on the chemical potential $\mu$, but they must be solved
under the constraint of a given fermion density which does depend on the chemical potential. This point will be discussed
in Section 6.

For  $U_0=1$, $ M=M^{\dagger}$ and time-independent  external fields, we can look for time-independent solutions, so that the previous equations become
\begin{eqnarray}
{\mathcal F} &= &N + e^{-2M} {\mathcal F} \, {\mathcal F}_{N}^{-1}
 e^{-2M}
 \nonumber\\
 {\mathcal F}^{\dagger} &=& N^\dag +e^{-2 M}\big({\mathcal F}_{N}^\dag\big)^{-1} {\mathcal F}^{\dagger} e^{-2 M}\, .
\end{eqnarray}
Now we make an ansatz  for the solutions of the form
\begin{equation}
\overline{{\mathcal F}}= NA\,.
\end{equation}
The matrix $A$ satisfies a quadratic equation. Assuming the commutation relations
\be
[N, M]  = 0
\ee
and
\be
[N^{\dagger}N, A] =0
\ee
we can solve for $A$ in a basis in which it is diagonal. We have then in general two solutions for each diagonal element. This multiplicity of solutions for $A$ can be related to the well known ambiguity in the Foldy-Wouthuysen transformation~\cite{EK}.
In the second quantization formalism this ambiguity can be solved by requiring that the vacuum energy be minimal. As a consequence  we find
\begin{equation}
A = (2N^{\dagger}N)^{-1}  \left[-Y
  + \sqrt{ Y^2 + 4  N^{\dagger}N} \right] 
\end{equation}
where
\begin{equation}
Y=1- N^{\dagger}N - e^{-4M}\,.
\end{equation}
For Kogut-Susskind fermions  $M=0$ and
\begin{equation}
N^{\dagger} N = 4 \, H^2  \label{pippo}
\end{equation}
where $H$ is the lattice Hamiltonian 
\begin{equation}
H^2 =  m ^2 -  \Delta
\end{equation}
with
\begin{equation}
\Delta = \frac{1}{4}   \sum_{i=1,3} \left( T_i^{(+)}+ T_i^{(-)} -2 \right) \,.   \label{laplascian}
\end{equation}
Then~\cite{Cara}
\begin{equation}\label{GSbosoncompo}
A =  (2\, H)^{-1} \left (H+\sqrt{1+H^2} \right)
\end{equation}
and using this expresson we derive that
\begin{align*}
\overline{\cal H} \, = & \,  e^{s\mu}\, H\, \left( \sqrt{1 + H^2} - H \right) \\
\giro{\overline{\cal H}} \,  = & \, e^{-s\mu}\, H\, \left( \sqrt{1 + H^2} - H \right)
\end{align*}
so that in the formal continuum limit
\be
\overline{\cal H} \, \approx  \, \giro{\overline{\cal H}} \, \approx \, H =  \sqrt{ m^2 - \bigtriangleup } 
\ee
both approach the same value.


\section{Chiral and vector symmetries}

Now we want to verify that the condensation generated by the factorization of the transfer
matrix does not break any symmetry, namely the vacuum
$ |\mathcal{F} \rangle$ remains invariant.
For this purpose we can consider only the Kogut-Susskind formulation in which  there is a residual chiral invariance. In this case chiral and vector transformations are 
\begin{eqnarray}
\psi' &=&\exp(i \theta_5 \gamma_5\otimes t_5 + i \theta)\psi
\nonumber\\
\overline{\psi }'&=&\exp(i \theta_5 \gamma_5\otimes t_5 -i \theta)\psi
\end{eqnarray}
so that
\begin{eqnarray}
u' &=& \exp(i \theta_5 \gamma_5\otimes t_5 + i \theta)u
\nonumber\\ 
{v^{\dagger}} '&=&\exp(i \theta_5 \gamma_5\otimes t_5 +i \theta) v^{\dagger} \,.
\end{eqnarray}
The operator which implements chiral transformations is
\be
\Lambda = \exp( i \theta_5 \gamma_5\otimes t_5 )
\ee
and therefore
\be
\Lambda^{\dagger} N \Lambda = - 2 i m \gamma_0 \otimes  I  \exp (2 i \theta_5 \gamma_5 \otimes t_5)\,.
\ee
Given the form of $\mathcal{F}$ for massive spinors
\be
\Lambda|\mathcal{F} \rangle  \neq |\mathcal{F} \rangle \,.
\ee
Neither the original theory nor the new vacuum are chiral invariant.

When we consider  the case of massless spinors, instead, $\Lambda$ commutes with $N$, so that
\be
\Lambda|\mathcal{F} \rangle  = |\mathcal{F} \rangle \,.
\ee
Both the original theory and the new vacuum are chiral invariant.

This particular case of massless spinors allows a factorization of the transfer matrix by using  the chiral components
\be
\chi_{\pm}:= \Pi_{\pm} \psi  \qquad \overline{ \chi}_{\pm}:= \overline{\psi} \, \Pi_{\pm} 
\ee
where the chiral projectors are defined as
\be
\Pi_{\pm} := \frac{ 1}{ 2} ( 1 \pm \gamma_5\otimes t_5 )\, .
\ee
The chiral components transform according to
\be
\chi_{\pm} '= \exp ( \pm i \theta_5 + i \theta)\chi_{\pm} 
\qquad \overline{ \chi}_{\pm}' =  \overline{ \chi}_{\pm}
\exp( \mp i \theta_5 - i \theta) \,.
\ee

\section{Connection with the Foldy-Wouthuysen\\ transformation}

There must be a relation between the Bogoliubov transformation, which factorizes the transfer matrix,
and the Foldy-Wouthuysen one which separates particles from antiparticles in the Dirac Hamiltonian. In
particular one might expect that in the formal continuum limit they should coincide.

To clarify this point we study the formal continuum limit of the expression the Bogoliubov transformation takes at the saddle
point. We observe that with both Wilson and Kogut-Susskind regularizations the following relations hold true
\be
{\overline {\mathcal F} } = {\overline {\mathcal F} }^{\dagger}
\ee
and
\be
P_{\pm} {\overline {\mathcal F} }  = {\overline {\mathcal F} } \, P_{\mp} 
\ee
so that $ {\overline {\mathcal F} }$ is odd under the transformation $P_+ - P_-$
\be
\left( P_+ - P_-\right)\,  {\overline {\mathcal F} }\, \left( P_+ - P_-\right)  = -  {\overline {\mathcal F} }\, .
\ee
Then in both cases the Bogoliubov transformation at the saddle point is
\be \psi' = R^{\frac{1}{ 2}} \left[ 1 + {\overline {\mathcal F} } \left( P_{+} - P_{-} \right) \right] \psi\,.
\ee
With Wilson regularization
\be
{\overline {\mathcal F} }\approx \frac{ i \sigma \cdot \nabla }{  m +\sqrt{ m^2 - \bigtriangleup}} = - \gamma_0 \,\frac{\vec{\gamma}\cdot\vec{\nabla}}{  m +\sqrt{ m^2 - \bigtriangleup}} 
\ee
so that
\be
R \approx \frac{m + \sqrt{ m^2 - \bigtriangleup} }{ 2\sqrt{ m^2 - \bigtriangleup} }\,.
\ee
We used  the $\vec{\gamma}$-matrices in terms of the Pauli matrices using
a convention different from that of L\"uscher~\cite{Lusc}: our $\vec{\gamma}$-matrices have opposite sign
\be
\gamma_0 = \left(\begin{array}{cc}1 & 0 \\0 & -1\end{array}\right)
 \qquad \vec{\gamma} = \left(\begin{array}{cc}0 & - i \vec{\sigma} \\ i \vec{\sigma} & 0\end{array}\right)\,.
\ee
It easy to check that the transfer matrix becomes the exponential of minus the continuum Hamiltonian times
the temporal lattice spacing
\be
{\hat {\mathcal T}} \approx \exp\left[ - a \,{\overline {\hat \psi}} \,(m  +  \vec{\gamma}\cdot\vec{\nabla}) {\hat \psi}\right]
\ee
and the Bogoliubov transformation coincides with the Foldy-Wouthuysen transformation for wave functions
\be
\psi' \approx \left[ 2 \sqrt{ m^2 - \bigtriangleup } \,(  \sqrt{ m^2 - \bigtriangleup } +m)\right]^{-1} \left[  \sqrt{ m^2 - \bigtriangleup }+m +  \vec{\gamma}\cdot\vec{\nabla} \right]  \, \psi \,.
\ee
The transformed Hamiltonian
\be
{\overline {\hat \psi'}} \,\left(m  +  \vec{\gamma}\cdot\vec{\nabla}\right)\, {\hat \psi}' = {\overline {\hat \psi'}} \, \gamma_0  \sqrt{ m^2 - \bigtriangleup } \, {\hat \psi}' 
\ee
is free of fermion-antifermion mixing and the states of positive/negative energy are those which propagate forward/
backward in time.

For Kogut-Susskind fermions
\be
{\overline {\mathcal F} }\approx - \gamma_0\, \frac{ m + \vec{\gamma}\cdot\vec{\nabla} }{ \sqrt{ m^2 - \bigtriangleup}}
\ee
so that 
\be
R \approx 2\,.
\ee
The transfer matrix is  approximately the exponential of minus the continuum Hamiltonian times the 
temporal block lattice spacing
\be
{\hat {\mathcal T}} \approx \exp \left[ - 2 a \,{\overline {\hat \psi}} \,(m  +  \vec{\gamma}\cdot\vec{\nabla}) {\hat \psi}\right] 
\ee
and the Bogoliubov transformation 
\be
\psi' \approx \left[ 2  \sqrt{ m^2 - \bigtriangleup } \right]^{-1} \left[  \sqrt{ m^2 - \bigtriangleup } +  \gamma_5 \otimes t_5 t_0 \, (m + \vec{\gamma}\cdot\vec{\nabla} ) \right]  \, \psi
\ee
does not have the form of the standard Foldy-Wouthuysen transformation. The
 transformed Hamiltonian in this case is 
\be
{\overline {\hat \psi'}} \,\left(m  +  \vec{\gamma}\cdot\vec{\nabla}\right)\, {\hat \psi}' =   {\overline {\hat \psi'}} \, \gamma_0\gamma_5 \otimes t_0 t_5 
 \sqrt{ m^2 - \bigtriangleup } \, {\hat \psi}' \,.
\ee
This is also free of fermion-antifermion mixing, and the states of positive (resp. negative) energy are those which 
propagate forward (resp. backward) in time on the lattice.

We remark that, in both regularizations, $ {\overline {\mathcal F} }$, in the continuum limit, is proportional to the part of the Hamiltonian odd under $P_+-P_-$\, .

\section{Finite temperature}

In this Section we shall consider the consequences of the identification of the physical modes in the case in which the temperature is finite.
To this end we first perform the integration over the quasi-particles fields getting the
effective action
\begin{multline}
   S_{\mbox{eff}}(\mathcal{F}) =  - \sum_{t=0}^{L_0/s-1} \tr_+ \ln \left(\mt \,U_{0,t}\,\me_{t+1,t} \right) \\
-   \Tr_+ \ln \left[\left( {\cal H}_t -\nabla_t  \right) T_0^{(+)}\right] -\Tr_-\ln\left[T_0^{(-)} \left( \giro{{\cal H}}_t - \giro{\nabla}_t \right)\right]
\end{multline}
where at variance with  ``$\tr$"  symbol  ``$\Tr$" is the trace over all fermion quantum numbers and   time labels.
 
For time-independent bosonic fields and for $U_{0,\,t}=\bone$ the effective action  
can be computed by using the saddle point solution $\overline{\cal F}$, which does not depend on time and thus on temperature, and we get
\begin{multline}
  S_{\mbox{eff}}(\overline{\cal F}) = -  \frac {L_0}{s} \tr_+ \ln \left(\mc \me\right)  \\
 -\Tr_+\ln\left[1-e^{s\mu}(\mc \me)^{-1}  T_0^{(+)}\right]-   \Tr_- \ln \left[1- e^{-s\mu}(\mc\me)^{-1}  T_0^{(-)}\right].
\end{multline}
After time-Fourier transformation
\begin{multline}
  S_{\mbox{eff}}(\overline{\cal F}) =  - \frac{L_0}{s}\,  \tr_+ \ln \left(\mc \me\right) \\
    -  \sum_{n=0}^{L_0/s-1} \,\tr_+ \,\left[
	\ln\left(1-e^{s\mu}( \mc \me )^{-1}\, e^{i \omega_n}\right) + \ln \left(1-e^{-s\mu} (\mc\me)^{-1}\, e^{-i \omega_n}\right)\right] 
\end{multline}
where
\be
\omega_n=\frac{s \pi}{L_0}(2n+1)\,,
\ee
with $n=0,\dots, L_0/{s}-1$, are the Matsubara frequencies. 
The corresponding sum, performed in Appendix~\ref{Matsubara}, gives 
\begin{multline}
S_{\mbox{eff}}(\overline{\cal F}) =  \,\tr_+ \left\{ \vphantom{\left(\giro{\overline{\cal H}}\right)^{-\frac{L_0}{s}}}   \frac{L_0}{s}\,   \ln\,\left( 1- e^{-s\mu}s\,\overline{\cal H}
 \vphantom{\left(\giro{\overline{\cal H}}\right)} \right)  \right. \\
- \left.  \ln \left[1+ e^{L_0\mu}\left(1 -e^{-s\mu} s\,\overline{\cal H} \vphantom{\giro{\overline{\cal H}}}\right)^{\frac{L_0}{s}} \right]
-    \ln\left[ 1+ e^{-L_0\mu} \left( 1- e^{s\mu}s\giro{\overline{\cal H} }\right)^{\frac{L_0}{s}}\right] \right\}\,.
\end{multline}
In the last equations we used the invariance of the trace under the involution (\ref{involution}) and the relations 
\begin{equation}
\left(\mc\me\right)^{-1}=1-e^{-s\mu}s\,\overline{\cal H} = 1-e^{s\mu}s\giro{\overline{\cal H}} .
\end{equation}
In the continuum limit
\be
S_{\mbox{eff}}(\overline{\cal F}) \approx  - \,\tr_+ \left[ L_0\, H + \ln \left( 1 + e^{- L_0 (H-\mu)} \right) +
\ln \left( 1 + e^{- L_0 (H+\mu)} \right) \right]
\ee
we recognize the contribution from the condensate energy and those from the Fermi statistics of the quasi-particles and quasi-antiparticles.

For $L_0\to\infty$ we recover the zero temperature result 
\begin{multline}
\lim_{L_0\to\infty}\frac{1}{L_0}  S_{\mbox{eff}}(\overline{\cal F}) = \,\frac{1}{s}\,\tr_+\left[\ln\, \left(1-e^{-s\mu}s\,\overline{{\cal H}}\vphantom{\left(\giro{\overline{\cal H}} \right)}   \right)  \right. \\
	-   \left. \theta\left(e^{s\mu}  -  1- s\,\overline{\cal H} \vphantom{\left(\giro{\overline{\cal H}}\right)}\right)\, \ln \left(e^{ s\mu}-s\,\overline{\cal H}  \vphantom{\left(\giro{\overline{\cal H}}\right)}\right) - \theta\left(e^{- s\mu} -  1- s\giro{\overline{\cal H}} \right)\, \ln \left(e^{- s\mu}-s\giro{\overline{\cal H}}\right)  \right] 
\end{multline}
which reduces to the one obtained in~\cite{Cara} by neglecting  the contribution of quasi-particles, and to the one obtained in~\cite{Palub} by neglecting only the contribution from antiquarks, which is given by the last term, an approximation  justified  for low values of temperature. 
Indeed, we can see the effect of the selection of a given number of fermions, by using the constraint
\be
- \frac{1}{L_0}\, \frac{\partial }{ \partial \mu}\, S_{\mbox {eff}}(\overline{\cal F}) = n_F 
\ee
we get at zero temperature
\be
\tr \,  \theta \left ( e^{s \mu} - 1 - s \, {\overline {\mathcal H}} \vphantom{ \giro{\overline  {\mathcal H}} } \right) - \tr  \, \theta
 \left( e^{ -s \mu} - 1 - s \, \giro{\overline  {\mathcal H}} \right)= n_F\, .
\ee
For non-negative $\mu$, the second $\theta$-function is always zero, meaning that quasi-antiparticles do not contribute.  Let us call $ {\overline \sigma}$ the smallest  mass in $H$. Near the continuum limit, for $\mu < {\overline \sigma}$,  $n_F=0$. For $\mu > {\overline \sigma}$, quasi-fermions occupy the states from zero energy up to a maximum energy depending on the fermion number $n_F$. 
The effect of a finite fermion density is to deplete the condensate, namely to reduce the number of fermionic states in the condensed boson structure function, without 
altering its form for the remaining states.

\section{Conclusion}

We have shown that the transfer matrix of relativistic field theories of fermions, whose 
Lagrangian can be written in a form quadratic in the fermion fields, can be factorized whenever
a certain classical equation has  solutions. In general there exists a multiplicity of solutions which can be restricted by
the requirement that the vacuum energy be minimal. 

Restricting  our analysis to the coupling
of fermions with time independent bosonic fields we have determined such solutions. They imply condensation of a composite boson which does not break any symmetry. 

Our main interest is in connection with our approach to bosonization, which can be used when the Fermi system is dominated at low energy by  excitations
which can be described in terms of bosonic composites. In such a case bosonization is achieved by
performing  independent
Bogoliubov transformations at each time slice, and associating the time dependent parameters 
of the transformations to  dynamical bosonic fields. The results of the present paper can then
become relevant when a saddle point expansion exists for such a bosonized system. In the presence
of spontaneous breaking of a symmetry, the  structure function iof the condensed bosons is in general a superposition of a symmetry breaking and a symmetry conserving
state, whose relative weight depends on temperature and chemical potential.

We have also shown that factorization of the transfer matrix  is related to the elimination of fermion-antifermion mixing in the Dirac Hamiltonian. Again there are many solutions to this latter probem\cite{EK}, which can be restricted considering it in second quantization and choosing the ones of minimal energy. It would be very interesting to explore the possibility 
of determining the Bogoliubov transformations corresponding to the   Foldy-Wouthuysen ones for fermions interacting with 
gauge fields.

\section*{Acknowledgements}

The work by F.~P.~ has been partially 
  supported by EEC under the contract MRTN-CT-2004-005104.


\appendix

\section{The matrices $M,N$ of the transfer matrix\label{transfer}}

In this Appendix we report the expressions of the matrices $M, N$ appearing in the definition of the transfer matrix for the
Kogut-Susskind and  the Wilson regularization. Their   common feature is that they  depend only on the spatial link
variables.

\subsection{ Kogut-Susskind's regularization }

Kogut-Susskind fermions in the flavor basis are defined on hypercubes whose sides are  twice the basic lattice spacing. 
While in the text intrinsic quantum numbers and spatial coordinates were comprehensively  represented by one
index $i$, here we distinguish the spinorial index $\alpha=\{1,\ldots,4\}$, the taste index 
$a=\{1,\ldots,4\}$ and the flavour index i=$\{1,...,N_f\}$, while $x=\{t,x_1,\ldots,x_3\}$ is a 4-vector of {\em even} 
integer coordinates ranging in the intervals $[0, L_t-1]$ for the time component and $[0, L_s-1]$ for each of the spatial components.
We distinguish summations over  basic lattice and hypercubes according to
\begin{equation}
\sum_x{}^\prime := 2^d \sum_x \,.
\end{equation}
The projection operators over fermions-antifermion states are
\begin{equation}
P_{\pm} = { \frac{1}{2} } ( \bone \otimes \bone \mp \gamma_0 \gamma_5 \otimes t_5 t_0 ) \, .
\end{equation}
In the presence  of gauge fields, neglecting an irrelevant constant,  $M=0$, 
while $N$ is~\cite{Palu1}
\be
N =  -2  \Big\{  m\,  \gamma_0  \otimes \bone  
  +{ \sum_{j=1}^3}  \gamma_0  \gamma_j   \otimes \bone\left[P^{(-)}_j\nabla_j^{(+)}+ P^{(+)}_j\nabla_j^{(-)}\right]  \vphantom{\sum_{j=1}^3} \Big\}
\ee
where
\begin{eqnarray}
 \nabla_j^{(+)} & = & \frac{1}{2} \left( U_j \,T^{(+)}_j  - 1 \right) \\
 \nabla_j^{(-)} & = &  \frac{1}{2} \left( 1- T^{(-)}_j  U_j^{\dagger}\right)
 \end{eqnarray}
 are the lattice covariant derivative and
\begin{equation}
P_j^{(\pm)} = { \frac{1}{2} } ( \bone \otimes \bone \mp \gamma_j \gamma_5 \otimes t_5 t_0 ) \, .
\end{equation}
The eigenvalues of $H^2  $ are  the fermion energies
\begin{equation}
E_q^2 =  m^2  + \tilde{p}^2 \, , \label{energy}
\end{equation}
where  momentum component $\tilde{p}^2_i $ is 
\begin{equation}
\tilde{p}^2_i = \frac{1}{2}  ( 1  - \cos 2\, p_i) \label{momentum}
\end{equation}
and 
\begin{equation}
\tilde{p}^2 = \sum_{i=1}^3 \tilde{p}^2_i
\end{equation}

\subsection{ Wilson's regularization }

The projection operators over fermions-antifermions are
\begin{equation}
P_{\pm} = { \frac{1}{2} } ( 1 \pm \gamma_0 ) \,.
\end{equation}

The matrices $M, N$ are
\begin{eqnarray}
M&=&  { \frac{1}{2} } \ln \left( \frac{B}{2K} \right)\\
N &= & 2 K \, B^{- \frac{1}{2} }\,c \, B^{-\frac{1}{2} } \,,
\end{eqnarray}
where
\begin{equation}
B = 1 -  K \sum_{j=1}^3 \left (U_j T^{(+)}_j + T^{(-)}_j U^\dagger_j  \right) 
 \end{equation}
$K$ is the hopping parameter and
\begin{equation}
c = { \frac{1}{2} }  \sum_{j=1}^3 i \left ( U_j  \, T^{(+)}_j - \, T^{(-)}_j U^\dagger_j  \right)\,
\sigma_j \,.
\end{equation}

\section{The unitary operator of the Bogoliubov transformation \label{transformation} }

In the framework of the BCS theory, the relation between the BCS wave function and the
Bogoliubov transformation was shown by Yosida~\cite{Yosida,Blatt} who constructed
a unitary operator which transforms fermion creation-destruction operators into 
quasi-particle operators. It can be of some interest to construct  the corresponding
operator for our generalized Bogoliubov transformation.

We write the operator in the form
\begin{equation}\label{defhatUbog}
 \hat{\mathscr U}:= e^{\hat S}
\end{equation}
where
\be
    {\hat S} :=\uh^ \dag X ^ \dag \vh^ \dag-\vh X \uh\,.
\ee
First we determine the action of the the operator ${\hat S} $ on creation and destruction operators
\begin{align}
\left[\hat S ,\uh \right] &=  -X ^ \dag \vh^\dag ,&
\left[\hat S ,\vh^\dag \right]& =   X \uh \\
\left[\hat S,\left[\hat S,\uh \right]  \right]& =  -X^ \dag X \uh ,  
&\left[\hat S,\left[\hat S,\vh^\dag \right]\right]&=  -X X^\dag \vh ^\dag \,.
\end{align}
 Using the identity
\begin{equation}\label{algebra1}
e^{A} B e^{-A}=\sum_{n=0}^\infty \frac{1}{n!}[A,[A,\cdots[A,B]\cdots]]
\end{equation}
for $e^{A}=e^{\hat S}, B=\uh, \,\vh^{\dagger} $ we find the transformations of creation destruction operators
\begin{align}\label{bogolibovtrasmorm}
e^{\hat S}  \uh\, e^{-\hat S} &=\sum_{n=0}^\infty\frac{(-X^\dag X )^n}{(2n)!}
		\left[\uh-\frac{1}{2n+1} X^\dag \vh\right]\\
		&=\cos\left(\sqrt{X^\dag X}\right) \left[\uh-\tg\left(\sqrt{X^\dag X}\right)\left( X^{\dag} X\right)^{-\meta}X^{\dag}\vh^\dag\right]\,,\\
e^{S} \vh^\dag e^{-S}&=\sum_{n=0}^\infty \frac{(-X X^\dag )^n}{(2n)!} \left[ \vh^\dag +\frac{1}{2n+1} X
			\uh\right] \\
		&=\cos\left(\sqrt{X X^\dag}\right) \left[\vh^\dag+\tg \left(\sqrt{X X^\dag}\right)
		\left( X^{\dag} X\right)^{-\meta}X\uh \right]\,.
\end{align}
They coincide with the transformations (\ref{Bogoliubov}) after the identification
\be
\frac{ 1}{ \sqrt{X^{\dagger}X} }\, \tg\sqrt{X^{\dagger}X} \,\,X^{\dagger}= \mathcal{F}^{\dagger} \,.\label{identification}
\ee
Next we determine the transformation of the vacuum. To this end it is convenient to write
the operator $\hat{S}$  in the form
\begin{align}
 \hat{S}=\sum_{i}\uh^ \dag_{i} \hat{\Lambda}_{i}^ \dag-{\hat{\Lambda}}_{i} \uh_ {i}=\sum_{i}\hat{\sigma}_{i}\,,&& 
\end{align}
where
\begin{align}
   {\hat{\Lambda}}_{i}:= \sum_j\vh_j  X_{ji} &&
\hat{\sigma}_{i}:=(\uh^\dag_{i}\hat{\Lambda}_{i}^ \dag-\hat{\Lambda}_{i}\uh_i)\,.
\end{align}
It follows that $\left\{\uh_{j}\,,\,{\hat\Lambda}_{i}\right\}=0$ and 
\begin{align}\label{evparP2}
  \big[\sigma_{i},\sigma_{j}\big]=\uh_j ^\dag \left(X^\dag\cdot X\right)_{i,j}\uh_i-\uh_i ^\dag \left(X^\dag\cdot X \right)_{i,j}\uh_j\,.
\end{align}
The matrix  $X^\dag X $ can be diagonlized by the transformation $\uh\mapsto (O \uh)_k=:{\tilde u}_k$. The new operators satisfy the relation $\left(O X^\dag X O^{-1}\right)_{k,k'}=\delta_{k,k'}(\tilde{X}^\dag \tilde{X})_{k}$, so that the commutator in the above equation vanishes. Therefore
\begin{equation}\label{evparP3}
  e^{\hat{S}}=\prod_{k} e^{\,\tilde{\sigma}_{k} }
\end{equation}
and writing ${\tilde \sigma}_{i}:=(\tilde {u}^\dag_{i}\tilde{\Lambda}_{i}^ \dag-{\tilde \Lambda}_{i}{\tilde u}_i)\,,$ ${\tilde \Lambda}_{i}=(\Lambda\cdot O^{-1})_i$ we have
\begin{align}
 \tilde{\sigma}_{k}^{2}&=-\left( \tilde{\Lambda}_{k}^\dag\tilde{\Lambda}_{k}^{\vphantom{\dagger}}
  \tilde{u}_{k}^ \dag\tilde{u}_{k}^{\vphantom{\dagger}}+\tilde{\Lambda}_{k}^{\vphantom{\dagger}} \tilde{\Lambda}_{k}^\dag\ut_{k}^{\vphantom{\dagger}}\ut_{k}^ \dag\right)\\
\tilde{\sigma}_{k}^{2n}&=(-1)^n\left[\left(\tilde{\Lambda}_{k}^\dag\tilde{\Lambda}_{k}^{\vphantom{\dagger}}\right)^{n} \ut_{k}^ \dag\ut_{k}^{\vphantom{\dagger}}+\left(\tilde{\Lambda}_{k}^{\vphantom{\dagger}} \tilde{\Lambda}_{k}^ \dag\right)^{n}\ut_{k}^{\vphantom{\dagger}}\ut_{k}^ \dag\right]\\
 e^{\tilde{\sigma}_{k} }&=\sum_{n} \frac{(-1)^n}{(2n)!}\left[1+\frac{\tilde{\pezzo}_{k}}{2n+1}\right] 
      			\left[\left(\tilde{\Lambda}_{k}^{\dagger}\tilde{\Lambda}_{k}^{\vphantom{\dagger}}\right)^{n}\ut_{k}^\dag\ut_{k}^{\vphantom{\dagger}}+\left(\tilde{\Lambda}_{k}^{\vphantom{\dagger}} \tilde{\Lambda}_
 	 {k}^\dagger\right)^{n}\ut_{k}^{\vphantom{\dagger}}\ut_{k}^{\dagger}\right]\\
e^{\tilde{\pezzo}_{k} }\ket{0}&=\sum_{n}\left.\frac{(-1)^n}{(2n)!}\left[1+
  \frac{\tilde{\pezzo}_{k}}{2n+1} \right]\left(\tilde{\Lambda}_{k} ^{\vphantom{\dagger}}\tilde{\Lambda}_{k}^{\dagger}\right)^{n}	\ut_{k}^{\vphantom{\dagger}}\ut_{k}^ \dagger\ket{0}\,.\right.\label{newvacuum}
\end{align}
Observing that $\tilde{\Lambda}_{k}\tilde{\Lambda}_{k}^\dag\ket{0}=(\tilde{X}^\dag \tilde{X})_{k,k'}\ket{0}$ the last equation can be rewritten
\begin{align}
  e^{\tilde{\pezzo}_{k} }\ket{0}&= \cos \sqrt{{\tilde X}^\dag_k{\tilde X}_k^{\vphantom{\dagger}} }
          e^{\tg\sqrt{ {\tilde X }^\dag_k{\tilde X}_k ^{\vphantom{\dagger}}}  \,
             \left({\tilde X}_k^{\dag} {\tilde X_k^{\vphantom{\dagger}}}\right)^{-\meta}  \ut_k^\dag 
          \tilde{\Lambda}_{k}^\dagger } \ket{0}\label{evparP6c}\,.
\end{align}
Using this equation in Eq.\eqref{evparP3} we have
\be\label{evparP7}
\prod_k  e^{\tilde{ \sigma }_{k} } \ket{0}= \left( \prod_k \cos\sqrt{{\tilde X }^\dag_k{\tilde X }_k^{\vphantom
			{\dagger}} }  \right)\,
e^{\sum_k\tg\sqrt{{\tilde X  }^\dag_k{\tilde X  }_k^{\vphantom{\dagger}}}  \,     \left({\tilde X  }_k^{\dag} {\tilde X  _k^{\vphantom{\dagger}}}\right)^{-\meta}  {\ut}_k^\dag  \tilde{\Lambda}_{k}^\dag}\ket{0}
\ee
so that in the original basis
\begin{align}\label{evparP9}
e^{\hat{S}} \ket{0}=&\, \det_{\!+} \cos\sqrt{X^\dag X} 
  \, e^{\uh^\dag  \tg\left(\sqrt{X^\dag X } \right)\left( X^{\dag} X\right)^{-\meta} X^\dag{\vh }^\dag}\ket{0}\\
  =&\,\det_{\!+} \cos\sqrt{X^\dag X} \,\,e^{\uh^{\dagger}\fc^{\dagger}\vh^{\dagger } }\ket{0}
\end{align}
where we used the identification in Eq. \eqref{identification}

\section{Evaluation of the trace in the partition function \label{trace} }

In this section, we sketch the evaluation of the matrix element of the transfer matrix in the quasi-particles coherent state base Eq. \eqref{bogcoherent}, and the effective action in Eq. \eqref{Z}. During the computation we use the property that in the transfer matrices and in the quasi-particles coherent state, the fermionic operators appears always in the exponential; so we can evaluate the matrix element in the coherent state bases performing some Berezin integral.

The coherent states for particles and antiparticles are defined in Eq. \eqref{coherent}, we can write the identity by this state as follow 
\begin{align}\label{identity}
{\hat \Iddop}=\int{{\mathrm D}\omega^*\, {\mathrm D} \omega {\mathrm D} \varphi^*
\misuradif \varphi  \frac{\ket{\omega,\varphi}\bra{\omega,\varphi}}{\braket{\omega,\varphi}{\omega,\varphi}}}\,.
\end{align}
Then we insert ${\hat \Iddop}$ between each factor of the transfer matrix 
\begin{multline}
\bra{\alpha_t \beta_t ; \phi_t}\hat {T}_t^{\dag} \hat {V}_t e^{s\mu\hat B}\hat {T}_{t+1}\ket{ \alpha_{t+1} 				\beta_{t+1};\Fifield_{t+1}}= \\
=\bra{\alpha_t \beta_t ; \phi_t}\hat {T}_t^{\dag}\,{\hat \Iddop}\,  \hat {V}_t e^{s\mu \hat n}{\hat \Iddop}\hat {T}_{t+1}\ket{ \alpha_{t+1}\beta_{t+1} ;\Fifield_{t+1}}\,
\end{multline} 
and between the two factors of $\hat {T_t}^{\dag}$ and $\hat {T}_{t+1}^{\vphantom{\dag}}$. The presence of coherent state permit us to substituite every operators with the a Grassmann variable. So we can arrive at the value of the matrix element of transfer matrix  performing the Berezin integral respect the Grassmannian variables appearing in  ${\hat \Iddop}$. Some intermediate results are 
\begin{multline*}
\bra{\rho \sigma } e^{wNu} \ket{ \alpha \beta;\Fifield}=\bra{\rho \sigma } e^{wNu}\,{\hat \Iddop} \ket{ \alpha \beta;\Fifield}= \\
\det_{\! +}  {\cal F}_N^\dag
\, \exp \Big[ -\beta \fc \alpha+ b N (\fc_{N}^\dag)^{-1}  a+ \rho^{*} (\fc_{N}^\dag)^{-1}  a - b (\fc_{N}^\dag)^{-1} \sigma^{*}+ \rho^{*}\fc_N \fc^\dag \sigma^{*} \Big]
 \end{multline*}
 \be
\bra{\omega,\varphi}e^{ - \uh^\dag  A \uh-\vh^\dag B \vh
}\ket{\rho,\sigma}=\exp\left(\omega^*e^{-A}\rho
+\varphi^*e^{-B}\sigma\right)\,.
\ee
Where we defined $a:=\mc^{-\meta}\alpha$ and $b:=\beta \giro{R}{\hspace{-1mm}}^{-\frac{1}{2}}$\,. The integration over $\rho$ and $\sigma$ and their h.c. of the product of the matrix element, in the last equations, gives
\begin{align}
& \bra{ \omega \varphi} \hat V_T e^{\mu \hat n_B}\hat T_{t+1} \ket{\alpha \beta;\Fifield }
			= \det_{\!+}  \fc_{N\, t+1}^\dag \exp \Big[-\beta \fc_{t+1}\alpha+
 b N_{t+1}(\fc_{N\,t+1}^\dag)^{-1} a \nonumber \\
& \qquad +\omega^* e^{s\mu}  U_{0,\,t}e^{-M_{t+1}}(\fc_{N\,t+1}^{\dag})^{-1} a
 -b\, e^{-s\mu} (\giro{\fc}_{N\,t+1}^{\,\dag})^{-1}e^{-M_{t+1}}U_{0,\,t}^\dag \varphi^*+\nonumber \\
 & \qquad + \omega^* U_{0,\,t}e^{-M_{t+1}} (\fc_{N\,t+1}^\dag)^{-1}\fc^\dag_{t+1} e^{-M_{t+1}}U_{0,\,t}^\dag \varphi^*\Big]\,.
\end{align}
In the last equations we highlight the time index.
The last equation, its Hermitian adjoint and a integral over the complex variable $\omega$ and $\varphi$ permit us to write the partition function in path integral formalism;  the effective action can be split in two contribute $S_0 (\fc)$ defined in Eq. \eqref{bosonaction} and 
\be
 S_F(\alpha,\beta;\fc)= 
 \sum_{t=0}^{L_0/2-1}\Big[- a^*_t \Laa_t a_{t} + b_{t} \Lbb _t b^*_{t} 
	-  b_{t} \Iaa_t^{(2,1)} a_{t} - a^*_{t}\Iaa{}_t^{(1,2)} b^*_{t} \Big]
\ee
where
\begin{align}
\Laa_t \, := &\, - \mt+  \me^{-1}_{t+1,t}~e^{s\mu}~T^{(+)} \\
\Lbb_t \, := &\,-\nt+e^{-s\mu}~T^{(-)}~ \md_{t+1,t}^{-1} \, .
\end{align}
It can be put in a more transparent form as in Eq.~\eqref{SF}.

\section{Sum over the Matsubara frequencies \label{Matsubara} }

We want to prove the following equation
\be
\sum_{n=0}^{N-1}\ln \left(1-\frac{e^{i\omega_n}}{c} \right)=\ln \left(1+\frac{1}{c^{N}}\right)\,.
\ee
where
\be
\omega_n=\frac{ \pi}{N}(2n+1)\, .
\ee
Taking the exponential of the left hand side we get
\be
\prod_{n=0}^{N-1}\left(1 -\frac{e^{i\omega_n}}{c} \right)= 
	\frac{e^{i\pi}}{c^{N}} \prod_{n=0}^{N-1}\left(c \, e^{-i\frac{\pi}{N}}-e^{i\frac{2\pi}{N}n} \right)\,.
\ee	
Now we use the identity
\be
\prod_{n=0}^{N-1}\left(A -e^{i\frac{2\pi}{N}n} \right)=A^N-1
\ee
 (which an obvious consequence of the fact that the $e^{i\frac{2\pi}{N}n}$are the $N$-th roots of   $1$) valid for arbitrary $N$ and $A$, to get the final result 
\be
\prod_{n=0}^{N-1}\left(1 -\frac{e^{i\omega_n}}{c} \right)= 1+\frac{1}{c^{N}}\,.
\ee

\end{document}